\documentclass[reprint,amsmath,amssymb,aps,pra,]{revtex4-1}
\usepackage{graphicx}
\usepackage{dcolumn}
\usepackage{bm}
\usepackage{soul}
\usepackage{color}
\usepackage{float}

\usepackage[ruled]{algorithm2e}
\usepackage{algpseudocode}

\usepackage{amsmath}
\usepackage{amssymb,amsfonts}

\newtheorem{thm}{Theorem}

\usepackage{makecell}
\usepackage{enumitem}
\usepackage{pgf,pgfarrows,pgfnodes,pgfautomata,pgfheaps,pgfshade}
\usepackage{tikz}
\usetikzlibrary{shapes,arrows,automata}
\usetikzlibrary{positioning}
\usetikzlibrary{decorations.pathmorphing} 
\usetikzlibrary{matrix} 
\usetikzlibrary{calc} 
\usetikzlibrary{trees}
\usetikzlibrary{decorations.markings}
\usetikzlibrary{shapes.geometric, arrows}

\tikzstyle{block} = [draw,rectangle,text centered,thick,minimum height=2em,minimum width=2em]
\tikzstyle{sum} = [draw,circle,inner sep=0mm,minimum size=2mm]
\tikzstyle{process} = [rectangle, rounded corners, minimum width=3.5cm, minimum height=1cm, text centered, draw=black]
\tikzstyle{decision} = [diamond, minimum width=3cm, minimum height=1cm, text centered, draw=black]

\begin{document}

\preprint{APS/123-QED}

\title{Anonymous communication protocol over quantum networks}
\author{Beili Gong}
\author{Wei Cui}
\email{aucuiwei@scut.edu.cn}
\affiliation{School of Automation Science and Engineering, South China University of Technology, Guangzhou 510641, China
}

\date{\today}

\begin{abstract}
We propose a $W$ state-based protocol for anonymously transmitting quantum messages in a quantum network.
Different from the existing protocols [A. Unnikrishnan, et al., Phys. Rev. Lett. 122, 240501 (2019)], the proposed protocol can be effectively implemented in the network only equipped with quantum channels and regular broadcast channels.
Throughout the design procedure, we develop three sub-protocols using the $W$ state, including the quantum collision detection protocol and the quantum notification protocol.
Moreover, together with the conventional anonymous entanglement protocol, the whole anonymous communication protocol has been constructed.
Finally, we examine the correctness and security of the proposed quantum anonymous communication protocol. 
\end{abstract}

\maketitle
\section{Introduction}
Anonymity and privacy protection are indispensable in communication security.
Over the past few decades, a large number of anonymous communication protocols have been proposed \cite{chaum:1988,Reiter:1998,Lai:2019}.
However, these protocols are based on the assumption that most agents are honest, and their security are relied on the computational complexity.
These conditions make it difficult to resist an adversary with higher computational power.
Quantum information science has seen remarkable growth in the past five years  \cite{Andersen:2015,Hayashi:2017,Yin:2020,Watrous:2018,Parra:2020,Zhou2020}.
In particular, how to anonymously transmit  quantum messages over a quantum network has attracted extensive research interest due to its widely potential applications in anonymous ranking \cite{Huang:2014,Lin:2016,Wang:2020}, anonymous voting \cite{Vaccaro:2007,Wang:2016,Jiang:2020}, sealed-bid auctioning \cite{Naseri:2009,Shi:2019}, and so on.

The first quantum protocol for sending and receiving a quantum message anonymously through a quantum network was proposed by Christandl and Wehner \cite{Christandl:2005}.
They introduced a key concept called anonymous entanglement, i.e., creating an EPR pair between the sender and the receiver in an anonymous way, and then achieved quantum message transmission by quantum teleportation.
Based on anonymous entanglement, a number of anonymous communication protocols have been presented in recent years \cite{Wang:2010,Yang:2016,lipinska:2018,Unnikrishnan:2019,Zhou:2020}.
In general, a complete anonymous protocol for quantum message transmission mainly consists of four parts:  multiple-sender detection, receiver notification, anonymous entanglement, and quantum teleportation.
Corresponding to the first two parts, the frequently used solutions are the collision detection protocol and the notification protocol \cite{Broadbent:2007,Brassard:2007}.
Since the two classical sub-protocols are information-theoretically secure in the presence of an unlimited number of misbehaving agents with polynomial computing power, which makes the existing protocols unable to resist an adversary with higher computing power than polynomials.
Moreover, most of these protocols can only accomplish the task of anonymous communication in a quantum network with quantum channels, simultaneous broadcast channels, and pairwise-private channels.
Inspired by establishing anonymous entanglement between a sender and a receiver using the $W$ state in Ref.~\cite{lipinska:2018}, we focus on designing an anonymous communication protocol that is resistant to an adversary with unlimited computational power and is easily implemented in a quantum network.

In this manuscript, we demonstrate how to anonymously transmit quantum messages step by step, and present the corresponding sub-protocols, including a quantum collision detection protocol, a quantum notification protocol, an anonymous entanglement protocol, and an anonymous bit transmission protocol.
Based on the shared $W$ state and the private lists of the agents, the anonymous communication protocol is proposed.
The implementation of the proposed protocol in the quantum network only requires public quantum channels and regular (or nonsimultaneous) broadcast channels.
We give a detailed analysis of the protocol correctness and the protocol security.
Theorem~\ref{thm1} shows that the correctness of the proposed protocol can be guaranteed if the source and all agents are honest.
In a semiactive adversary scenario (an active adversary and a trusted source), the anonymity and the security of the proposed protocol are proved by Theorems~\ref{thm2} and~\ref{thm3}, respectively.

The manuscript is organized as follows.
We present a detailed quantum anonymous communication protocol in Sec.~\ref{The_Protocol}.
In Sec.~\ref{correctness_security}, we consider the protocol's correctness, as well as its anonymity and security with a semiactive adversary attack.
Finally, we summarize our conclusions in Sec.~\ref{Conclusion}.

\section{The Protocol}\label{The_Protocol}
The task of quantum anonymous communication protocol aims at anonymously transmitting an arbitrary quantum state from a sender to a receiver over a quantum network.
To define the task more precisely, consider a quantum network with $n$ agents, $P_1, P_2, \dots, P_n$, who
can perform local operations and measurements, and a trusted source which is used to generate the quantum states required for anonymous communication.
Differing from the most existing networks in literature, only two communication channels are involved in our network, namely the regular channel and the quantum channel.
The former is used by the agents to broadcast classical information, while the latter is used by the source to distribute quantum states.
Here, all channels are assumed to be secure and noise-free, and the agents may be honest or not.
Also, the trusted source can be played by any honest agent on the condition that the choice of the agent is independent of who the sender is \cite{Unnikrishnan:2019}.

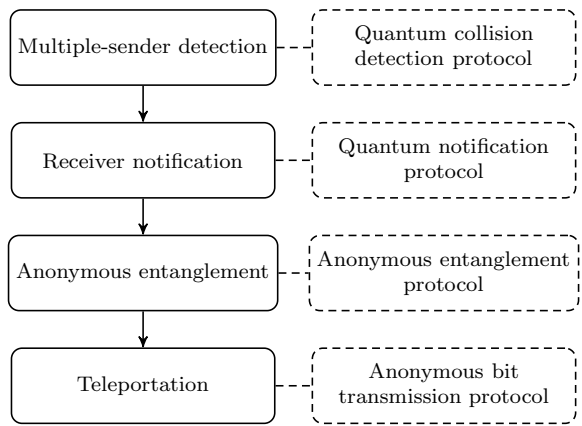
\begin{figure}[hbt]
\centering
\begin{tikzpicture}[auto, node distance=1.5cm, >=stealth', line width=0.6pt]
\footnotesize
\node [process] (pro1) {Multiple-sender detection};
\node [process, right of= pro1, densely dashed, xshift=2.5cm] (pro11) {\makecell[c]{Quantum collision \\ detection protocol}};
\node [process, below of=pro1] (pro2) {Receiver notification};
\node [process, right of= pro2, densely dashed, xshift=2.5cm] (pro21) {\makecell[c]{Quantum notification \\ protocol}};
\node [process, below of=pro2] (pro4) {Anonymous entanglement};
\node [process, right of=pro4, densely dashed, xshift=2.5cm] (pro41) {\makecell[c]{Anonymous entanglement \\protocol}};
\node [process, below of=pro4] (pro5) {Teleportation};
\node [process, right of=pro5, densely dashed, xshift=2.5cm] (pro51) {\makecell[c]{Anonymous bit \\transmission protocol}};

\draw[->](pro1) -- (pro2);
\draw[->](pro2) -- (pro4);
\draw[->](pro4) -- (pro5);

\draw[-, densely dashed](pro1) -- (pro11);
\draw[-, densely dashed](pro2) -- (pro21);
\draw[-, densely dashed](pro4) -- (pro41);
\draw[-, densely dashed](pro5) -- (pro51);
\end{tikzpicture}
\caption{Flowchart of constructing an anonymous communication protocol}
\label{Fig:Process}
\end{figure}

Under this network model, we design a novel quantum protocol for anonymous transmission, inspired by the well-known four-step anonymous protocol in Refs.~\cite{Brassard:2007,Wang:2010,lipinska:2018,Unnikrishnan:2019}.
The design procedure of the protocol is depicted in Fig.~\ref{Fig:Process}, where the corresponding sub-protocols are of quantum version and constructed by using $n$-partite $W$ states.
Additionally, the protocol requires a key ingredient that the agent $P_i,\,i\in [1,n]$ maintains a private list $\{r_i^1,r_i^2, \dots, r_i^j, \dots, r_i^n, a_i\}$, where $a_i= \oplus_{j = 1}^n {r_i^j}$ and $r_i^j \in \{0,1\}$.
The Boolean function $r_i^j$ would indicate the notification relationship between agents $P_i$ and $P_j$, i.e.,
$r_i^j(j\ne i)=1$ if $P_i$  is the sender and  $P_j$ is the receiver; otherwise, $r_i^j = 0$.
Also, we exclude the trivial cases that the sender or receiver are known a prior by the agents and that the sender and the receiver is the same agent.
That is, $r_i^j(j= i) \equiv 0$. 
Based on this prerequisite, the sub-protocols and the anonymous communication protocol are detailedly presented below.

\subsection{Quantum Collision Detection Protocol}
Transmission collision is inevitable when multiple agents want to be senders simultaneously in a shared network.
To ensure only one sender per transmission, we propose a quantum collision detection protocol by using the $W$ state and $\{a_i\}_{i=1}^n$ in the private lists, as shown in Protocol~\ref{collision_detection_protocol}.
The condition for Protocol~\ref{collision_detection_protocol} to pass is that both $y_i=0$ and $y_i=2$ would appear in $n$ experiments.
Otherwise, either no sender or multiple senders among the agents, which makes the protocol abort.

\begin{algorithm}[htb]
\caption{quantum collision detection protocol}
\label{collision_detection_protocol}
\begin{algorithmic}[1]
\Require
$n$-partite $W$ state, $\{a_i\}_{i=1}^n$.
\Ensure
Each agent gets $y_i$.
\State
The agents agree on $n$ orderings, with each ordering having a different last agent.
\State
For each ordering:
\begin{itemize}
\item A trusted source distributes the $n$-partite $W$ state to all agents.
\item The agent $P_i$ performs the operation as follow: if $a_i=1$, $P_i$ applies a $X(\sigma_x)$ operator to her qubit, otherwise she does nothing.
\item The agent $P_i$ measures her qubit in the computational basis $\{|0\rangle, |1\rangle\}$, and broadcasts the measurement outcome $z_i$ according to the current ordering.
\item The value $z= \sum\limits_{i = 1}^n {z_i}$ is computed, which equals $y_i$.
\end{itemize}
\State
If $y_i=0$ and $y_i=2$ occur in $n$ experiments, it indicates that only one sender exists in the quantum network; otherwise, the protocol aborts.
\end{algorithmic}
\end{algorithm}

\subsection{Quantum Notification Protocol}
After passing the proposed collision detection protocol, i.e., there is a unique sender among the agents,
how the sender anonymously notifies an agent of being the receiver becomes the next priority \cite{Broadbent:2007,Brassard:2007}.
Based on the shared $W$ state and the lists of the agents $\{r_i^j\}_{i,j=1}^n$, the quantum notification protocol is given in Protocol~\ref{notification_protocol}.
After the Protocol 2 is executed, one and only one agent will be informed to be the receiver.

\begin{algorithm}[htb]
\caption{quantum notification protocol}
\label{notification_protocol}
\begin{algorithmic}[1]
\Require
$n$-partite $W$ state, each agent $P_i$ holds a list $\{r_i^j\}_{j=1}^n$.
\Ensure
The sender notifies the receiver in an anonymous way.
\State
For each agent $P_i$:
\begin{itemize}
\item A trusted source generates the $n$-partite $W$ state, and act on a random but even number of X operators before distributing the state.
\item Each agent $P_j$ performs $X$ operator to her qubit if the $i$th value $r_j^i$ in the list $\{r_j^i\}_{i=1}^n$ is $1$, otherwise nothing is done.
\item Each agent $P_j$ measures its qubit in the computational basis and obtains the measured outcome $Y_j$.
\item Other agents broadcast their measurement outcomes to agent $P_i$.
\item The value $\bar y_i= \oplus_{j = 1}^n {Y_j}$ is computed only by agent $P_i$.
\end{itemize}
\State If agent $P_i$ obtains $\bar y_i=0$, then she is the receiver.
\end{algorithmic}
\end{algorithm}

\subsection{Anonymous Entanglement Protocol}
Quantum teleportation is a technology of transporting any quantum state from one agent to another, even in the absence of a physical channel linking these two agents.
Quantum message transmission via quantum teleportation lies in constructing entanglement state between the sender and the receiver \cite{Furusawa:1998,Sherson:2006,Wolfgang:2014l}.
Through $n$ agents sharing a $W$ state, a protocol to establish anonymous entanglement between the sender and the receiver has been proposed in Ref.~\cite{lipinska:2018}.
However, their approach requires not only anonymous broadcast channels, but also private channels between agents for executing the classical veto protocol \cite{Broadbent:2007}.
We reconstruct the anonymous entanglement protocol based on the $W$ state and $\{b_i\}_{i=0}^n$, where $b_i = a_i \oplus \bar{y}_i$ is the parity of XOR-ing $a_i$ with the output $\bar y_i$ obtained from Protocol~\ref{notification_protocol}.
The execution of Protocol~\ref{entanglement_protocol} does not rely on any pairwise-private channel and anonymous broadcast channel since no classical protocol is involved.

\begin{algorithm}[htb]
\caption{anonymous entanglement protocol}
\label{entanglement_protocol}
\begin{algorithmic}[1]
\Require
$n$ agents share a $W$ state, $\{b_i\}_{i=1}^n$.
\Ensure
EPR pair shared between the sender and the receiver.
\State
A trusted source distributes the $n$-partite $W$ state to all agents.
\State
Each agent $P_i$ measures in the computational basis if $b_i=0$ and broadcasts her outcome $\hat{y_i}$; otherwise does not perform any measurement but broadcasts $\hat{y_i}=0$, simultaneously.
\State
The value $Z=\sum\limits_{i=1}^{n} \hat{y_i}$ is computed.
$Z=0$ means that the sender and the receiver share the EPR pair.
\end{algorithmic}
\end{algorithm}

\subsection{Anonymous Bit Transmission Protocol}
Once Protocol~\ref{entanglement_protocol} has perfectly created an EPR state between the sender and the receiver, the last ingredient for anonymous transmission is the anonymous bit transmission protocol, which is used to transmit a classical bit $m \in \{0,1\}$ anonymously, as shown in Protocol~\ref{anonymous_protocol1}, provided that there is a unique sender in each round of transmission.

\begin{algorithm}[htb]
\caption{anonymous bit transmission protocol}
\label{anonymous_protocol1}
\begin{algorithmic}[1]
\Require
$n$ agents share a $W$ state.
\Ensure
The sender anonymously transfers a bit $m\in\{0,1\}$.
\State
A trusted source distributes the $n$-partite $W$ state to all agents.
\State
If $m=1$, the sender performs $X$ operator to her qubit; if $m=0$, there is no operation.
\State
Each agent $P_i$:
\begin{itemize}
  \item Measures her qubit in the computational basis.
  \item Broadcasts the measured outcome.
  \item Counts the total number of $1'$s $k$.
\end{itemize}
\State
If $k$ is odd, she concludes $m=0$; otherwise $m=1$.
\end{algorithmic}
\end{algorithm}

\subsection{Anonymous Communication Protocol}
Based on the corresponding sub-protocols proposed in the previous subsections, the complete anonymous communication protocol for quantum message transmission is given in Protocol~\ref{anonymous_protocol2}.
\begin{algorithm}[htb]
\caption{anonymous communication protocol}
\label{anonymous_protocol2}
\begin{algorithmic}[1]
\Require
$n$-partite $W$ state, each agent $P_i$ holds a private list $\{r_i^1,r_i^2, \dots, r_i^j, \dots, r_i^n, a_i\}$.
\Ensure
The sender anonymously transmits a quantum state to the receiver.
\State
Collision detection.

To determine only one agent wants to be the sender, the agents run Protocol~\ref{collision_detection_protocol} according to $\{a_i\}_{i=1}^n$. If the outputs of the protocol occurs $0$ and $2$, continue.
\State
Notification.

Based on $\{r_i^j\}_{i,j=1}^n$, the agents run Protocol~\ref{notification_protocol}, where each agent obtains $\bar y_i$ and the receiver is notified.
Together with $a_i$, $b_i = a_i \bigoplus \bar y_i$ is obtained by agent $P_i$.
\State
Anonymous entanglement.

The agents run Protocol~\ref{entanglement_protocol} in the light of $\{b_i\}_{i=1}^n$.
If the output is $0$, then anonymous entanglement is established, else abort.

\State
Teleportation.

The sender teleports a quantum state to the receiver.
Classical messages $m_0, m_1$ associated with teleportation are sent through Protocol~\ref{anonymous_protocol1}.

\end{algorithmic}
\end{algorithm}

Note that the implementation of Protocol~\ref{anonymous_protocol2} is dependent on the private lists maintained by the agents, the $W$ states generated by the trusted source, the classical broadcast channel, and the quantum channel.
Additionally, it is worth mentioning that all the proposed sub-protocols are completely traceless, which is an intuitive and key feature of quantum protocols \cite{Broadbent:2007,Menicucci:2018}.
Particularly, the communicating pairs are untraceable in Protocol~\ref{anonymous_protocol2}.

\section{Correctness and Security}\label{correctness_security}
When discussing protocols in the context of quantum cryptograph, a key question is how to access the communication security.
Broadly speaking, the security refers to protection against attacks.
In the type of network under consideration, the attackers originate from the agents other than the honest communicating pairs.
When all the agents behave honestly, the security is sometimes referred to as correctness, which indicates a protocol's ability to achieve quantum message transmission anonymously \cite{Graaf:1998,Broadbent:2015,Wang:2010,lipinska:2018}.
When dishonest agents exist in the network, the security is twofold: anonymity and privacy \cite{Wehner:2004,Christandl:2005}.
By anonymity, it mean that the protocol is to hide the identities of the sender and the receiver from being attacked by dishonest agents. 
By privacy, the transmitted content is protected.
That is, the transmitted quantum state will not be obtained by dishonest agents.
In this section, we discuss the correctness of Protocol~\ref{anonymous_protocol2} when all agents are honest, and analyze the security of Protocol~\ref{anonymous_protocol2} in a semiactive attack scenario \cite{lipinska:2018}.


\subsection{Correctness}

The next result proofs the correctness of Protocol~\ref{anonymous_protocol2}.

\begin{thm}\label{thm1}
If $n$ agents act honestly and Protocol~\ref{anonymous_protocol2} does not abort, the task of anonymous transmission of quantum state is accomplished perfectly.
\end{thm}

\emph{Proof.}
According to the procedure of Protocol~\ref{anonymous_protocol2}, a step-by-step proof of the correctness is given, provided that all agents are honest.

In Step 1, all agents run Protocol~\ref{collision_detection_protocol}.
Initially, a trusted source generates a $W$ state and distributes it to the agents.
If one agent wants to be the sender, she performs $X$ operation on her qubit.
Then each agent measures its qubit in the computational basis $\{|0\rangle, |1\rangle\}$ and broadcasts the measured outcome.
Since $X|1\rangle=|0\rangle$ and $ X|0\rangle=|1\rangle$, the sum of all broadcasted results depends on the number of potential senders.
Especially, if there is a unique sender, the sum of the broadcasted result would be $0$ or $2$ in one experiment.
As a result, that the results $0$ and $2$ appear in $n$ experiments shows the existence of a unique sender among the agents, and indicates the correctness of Protocol~\ref{collision_detection_protocol}.
Otherwise, the protocol would be aborted.

In Step 2, the agents are notified one after another anonymously according to Protocol~\ref{notification_protocol}.
The initial state shared by the agents is obtained by applying an even but random number of $X$ operators on the $W$ state.
If the sender $P_i$ selects agent $P_j(j\ne i)$ as her unique receiver, the corresponding Boolean function $r_i^j =1$.
Then, agent $P_i$ performs $X$ operation only when notifying the agent $P_j$, and the parity, calculated by $P_j$, of the measured outcomes over the computational basis is even.
By sharing the initial state, however, the parity, obtained by each non-receiver agent, is odd.
The agents are informed by the parities they held, respectively, about if they are the receiver.
Then the correctness of Protocol~\ref{notification_protocol} is proven and the receiver knows her identity anonymously.

The analysis of the step 3 follows from the correctness of the anonymous entanglement protocol in Ref.~\cite{lipinska:2018}.
The only difference here is that instead of an anonymous broadcast channel, all agents broadcast their classical messages over the regular broadcast channels.
Also, the sender and the receiver can know from the broadcast results whether their anonymous entanglement has been successfully established.

Through the above three steps, an EPR pair is shared between the communicating pair.
Then the sender can perfectly transmit a quantum state to the receiver by quantum teleportation.
Note that the transmission of the measurement results of the sender depends on Protocol~\ref{anonymous_protocol1},
which is similar to Protocol~\ref{collision_detection_protocol}. 
Consequently, the correctness can be guaranteed in Step 4.
Based on the above discussions, Protocol~\ref{anonymous_protocol2} works correctly and the proof is completed.

%
\hfill $\Box$

In the honest implementation, an interesting phenomenon occurs when counting the broadcasted results of Protocol~\ref{collision_detection_protocol}:
if the sum of the broadcasted results is always $1$, then there is no sender in the network;
if both $i$ and $i+2$ occur as the sum of the broadcasted results in $n$ experiments, where $i\in[0,n-2]$, then there are $i+1$ senders;
if the sum of the broadcasted results is always $n-1$, then all agents are senders.
This indicates that the quantum collision detection protocol can be used to verify the number of senders in a transmission.

\subsection{Security}
As mentioned previously, it is necessary to guarantee anonymity as well as privacy when some agents behave dishonestly.
Semi-honest and malicious agents are usually considered in analyzing the security of a quantum protocol,
see Refs.~\cite{Wehner:2004,Bouda:2007,Goldreich:2009}.
In the semi-honest model, the honest-but-curious agents will take advantage of all messages their obtain.
Specifically, they can either read or copy the message as desired without affecting the execution of the protocol.
In the malicious model, the attackers can actively cheat from the original prescription of the protocol, such as that the cheaters can try to attain some information about the input of honest agents or tamper with the output of the protocol.
Typically, both models are neatly encapsulated by two central entities called a passive adversary and an active adversary, respectively \cite{Christandl:2005,Wang:2010}.

Clearly, Protocol~\ref{anonymous_protocol2} is secure to a passive adversary since all messages obtained by semi-honest agents are only the measurement outputs, which are random and published.
In our network model, the adversary in the malicious model is semiactive, as defined in Ref.~\cite{lipinska:2018}.
Thus, we mainly consider the case of existing a semiactive adversary, who can corrupt some agents, record and tamper with all the information gained by these malicious agents during executing the protocol. 
Also, the number and the computational power of the semiactive adversary are unlimited.
Like in related works \cite{Brassard:2007,lipinska:2018}, quantum attacks are not taken into consideration in this work.
The following results demonstrate that Protocol~\ref{anonymous_protocol2} guarantee both the anonymity of the communicating pair and the privacy of the quantum message in a semiactive adversary scenario.

\begin{thm}\label{thm2}
The identities of the sender and the receiver cannot be guessed in the semiactive adversary scenario, no matter how many agents the adversary controls except the sender and the receiver.
\end{thm}

\emph{Proof.}
The proof is divided into 4 steps.
In Step 1, each agent performs local operation and measurement, in sequence, based on the list it holds and the computational basis.
However, due to the attacks from malicious agents, their broadcasted results would be changed, which causes Protocol~\ref{collision_detection_protocol} to abort or pass.
In either case, no adversary obtains any information about the identity of the sender, since all broadcast results can only be used to infer whether there exists a sending conflict.
Thus, the anonymity of the sender is guaranteed regardless of how many agents are controlled by a semiactive adversary.

In Step 2, each agent maintains a list of the agents to notify based on Protocol~\ref{notification_protocol}.
The output of the protocol only privately indicates to each agent whether she is the receiver, without giving any other information, such as the number or the source of the notification.
If some agents are governed by a semiactive adversary, the worst case would be that the parity of broadcast results changes from even to odd or vice versa, which prevents the receiver from being notified or makes the sender aware of the presence of an adversary.
Nevertheless, it reveals no information on the identities of the sender and the receiver.
Therefore, the proposed quantum notification protocol is perfectly anonymous.

In Step 3, there are two possible attack scenarios: one is that the sender and the receiver share the EPR pair while the broadcast claims there is no entanglement between them, which results in terminating Protocol~\ref{entanglement_protocol}; the other is that the establishment of anonymous entanglement fails while the broadcast results erroneously show that the entanglement is established, which makes quantum message transmission via teleportation impossible even if Protocol~\ref{entanglement_protocol} was passed.
In either case, the anonymity of the sender and the receiver can be maintained.

Finally, in Step 4 the sender transmits a quantum state to the receiver via quantum teleportation.
During the process, the sender performs the Bell-state measurement and sends the measured outcomes by Protocol~\ref{anonymous_protocol1}.
Also, the receiver can obtain the transmitted state without breaking the anonymity.
In the semiactive adversary scenario, some malicious agents change their broadcast values, which only cause the receiver to obtain an incorrect state.
Consequently, the identities of the sender and the receiver are hidden from other agents.

Not only executing multiple collision detection and receiver notification, but also executing anonymous entanglement
and quantum teleportation do not reveal the identities of the sender and the receiver, even in a semiactive attack scenario.
Therefore, the anonymity of Protocol~\ref{anonymous_protocol2} is perfect.

\hfill $\Box$

\begin{thm}\label{thm3}
Suppose the sender and the receiver act honestly in Protocol~\ref{anonymous_protocol2}.
Then the semiactive adversary obtains no information about the quantum message, even there are some corrupted agents in the network.
\end{thm}

\emph{Proof.}
The privacy of Protocol~\ref{anonymous_protocol2} primarily involves entanglement establishment and teleportation.
If all agents are honest, the output of the anonymous entanglement protocol is zero, which means that the sender and the receiver share the EPR pair after executing Step 3.
Then the sender transmits a quantum state to the receiver via quantum teleportation i.e., executing Step 4.
While there are two types of attack scenarios in these two steps.

The first is that the semiactive adversary attacks by governing some agents to change their broadcasted results, which causes the sum of the broadcasted results to change from non-zero to zero or from zero to non-zero in Step 3.
In the former case, the anonymous entanglement between the sender and the receiver is mistakenly considered to be unestablished, which makes the proposed protocol abort and there is no quantum message leakage.
In the latter case, the result mistakenly shows the establishment of the EPR pair.
However, this only makes quantum message transmission via teleportation impossible, but not leak the message.
Additionally, it is still possible to be attacked by the adversary when the sender transmits the measurement results to the receiver according to Protocol~\ref{anonymous_protocol1}, even if they have shared the EPR pair.
In such case, tampering with the broadcast results leads to misoperation of the receiver such that the receiver obtains an incorrect message, without any information leakag.

The second is that the semiactive adversary stops some agents from performing quantum measurement and forces them to broadcast $0$ based on $\{b_i\}_{i=1}^n$ they held.
In this case, Step 3 of Protocol~\ref{anonymous_protocol1} can be passed, while the communicating pair and the unmeasured agents share the $W$ state.
For convenience, we consider a three-agent case, where one of them is a malicious agent.
The shared quantum state can be expressed as
\begin{equation*}
\begin{aligned}
|W\rangle_2 &= \frac{1}{\sqrt{3}}|100\rangle_{srm}+|010\rangle_{srm}+|001\rangle_{srm},
\end{aligned}
\end{equation*}
where $s$, $r$, ${m}$ stand for the sender, the receiver and the malicious agent, respectively.
Suppose the quantum state that the sender wants to transmit by applying quantum teleportation is $|\phi \rangle = \alpha |0\rangle_s +\beta |1\rangle_s $, where $\alpha, \beta$ are arbitrary complex numbers.
Then the joint state is given by
\begin{equation*}
\begin{aligned}
|\Phi \rangle_0 &= |\phi \rangle \otimes|W\rangle_2\\
&= \frac{1}{\sqrt{3}}\Big[\alpha |0\rangle_s(|100\rangle_{srm}+|010\rangle_{srm}+|001\rangle_{srm}) \\
&~~~~+ \beta |1\rangle_s (|100\rangle_{srm}+|010\rangle_{srm}+|001\rangle_{srm})\Big].
\end{aligned}
\end{equation*}
After the sender performs \emph{Controlled-NOT} gate on her qubits and then sends the first qubit through \emph{Hadamard} gate, it holds
\begin{equation*}
\begin{aligned}
|\Phi \rangle _1 &= \frac{1}{\sqrt{6}}\Big[\alpha(|0\rangle_s \!+\! |1\rangle_s) (|100\rangle_{srm}\!+\!|010\rangle_{srm}\!+\!|001\rangle_{srm}) \\
&~~~+ \beta (|0\rangle_s - |1\rangle_s)(|000\rangle_{srm}\!+\!|110\rangle_{srm}+|101\rangle_{srm})\Big].
\end{aligned}
\end{equation*}
By simple algebraic calculation, the shared state can be further written as
\begin{equation*}
\begin{aligned}
|\Phi \rangle _1 &= \frac{1}{\sqrt{6}}\Big\{
   |00\rangle_{ss}\big[(\alpha|1\rangle_r \!+\! \beta|0\rangle_r)|0\rangle_m\!+\!\alpha|0\rangle_r|1\rangle_m\big] \\
&~~~~+ |01\rangle_{ss}\big[(\alpha|0\rangle_r + \beta|1\rangle_r)|0\rangle_m+\beta|0\rangle_r|1\rangle_m\big] \\
&~~~~+ |10\rangle_{ss}\big[(\alpha|1\rangle_r - \beta|0\rangle_r)|0\rangle_m+\alpha|0\rangle_r|1\rangle_m\big] \\
&~~~~+ |11\rangle_{ss}\big[(\alpha|0\rangle_r - \beta|1\rangle_r)|0\rangle_m-\beta|0\rangle_r|1\rangle_m\big]\Big\}.
\end{aligned}
\end{equation*}
As a result, the measured results $m_0, m_1$ are taken values in $00$, $10$, $01$ and $11$ after the sender performs Bell-state measurement.
Note that $m_0, m_1$ are broadcasted anonymously to the receiver, according to Protocol~\ref{anonymous_protocol1}.
From the expression of $|\Phi \rangle _1$, the malicious agent only obtains two possible results by measuring over the computational basis: when the measurement outcome is $0$, it infers that the anonymous transmission of quantum state is successful; otherwise, the malicious agent only knows that the quantum state fails to be transmitted.
In both cases, the privacy of the quantum state can be guaranteed.
The above analysis can be extended to $n$-agents case with at most $(n-2)$ malicious agents.
Also, the privacy of the quantum message can not be affected even if malicious agents change the broadcasted results during the execution of Protocol~\ref{anonymous_protocol1}.
The proof is established.
\hfill $\Box$

Heretofore, we have given the proofs for the correctness, the full anonymity of the sender and the receiver, and the privacy of the transmitted quantum message.

\section{Conclusion}\label{Conclusion}
Taking the $W$ state as the only quantum resource, we have proposed an anonymous communication protocol for quantum message transmission in a quantum network.
The protocol is composed of four innovative sub-protocols, including the quantum collision detection protocol, the quantum notification protocol, the anonymous entanglement protocol, and the anonymous bit transmission protocol.
The completions of these sub-protocols only relies on quantum channels and regular broadcast channels, which reduces the complexity of physical requirement of the protocol in the quantum network, comparing with other existing protocols.
We  have also shown the correctness of the protocol and found, as an interesting by-product, that the number of senders can be obtained by the proposed collision detection protocol when all agents and the source are honest.
Finally, the security of the protocol, including the anonymity of the communicating pair and the privacy of the transmitted quantum message, has been illustrated in a semiactive adversary scenario.

\begin{acknowledgements}
This work was supported by the National Natural Science Foundation of China under Grant 61873317  and in part by the  Guangdong Basic and Applied Basic Research Foundation under Grant 2020A1515011375.
\end{acknowledgements}

\end{document}